\documentclass[sigconf,9pt]{acmart}

\usepackage{multirow}
\usepackage{subfigure}
\usepackage{bbding}
\usepackage{mathtools}

\usepackage{algorithm}
\usepackage{algpseudocode}
\algtext*{EndFor}
\algtext*{EndFunction}
\algtext*{Until}
\algtext*{EndIf}
\algtext*{EndWhile}
\usepackage{amsmath}
\usepackage{array}

\usepackage{geometry}

\AtBeginDocument{%
  \providecommand\BibTeX{{%
    \normalfont B\kern-0.5em{\scshape i\kern-0.25em b}\kern-0.8em\TeX}}}

\usepackage{tikz}
\usepackage{xcolor}
\newcommand*\circled[1]{\tikz[baseline=(char.base)]{
            \node[shape=circle,fill,inner sep=1pt,scale=0.8] (char) {\textcolor{white}{#1}};}}

\copyrightyear{2025} 
\acmYear{2025} 
\setcopyright{acmlicensed}\acmConference[ASPDAC '25]{30th Asia and South
Pacific Design Automation Conference}{January 20--23, 2025}{Tokyo, Japan}
\acmBooktitle{30th Asia and South Pacific Design Automation Conference
(ASPDAC '25), January 20--23, 2025, Tokyo, Japan}
\acmDOI{10.1145/3658617.3697658}
\acmISBN{979-8-4007-0635-6/25/01}

\begin{document}








\begin{abstract}

Point cloud is an important data structure for a wide range of applications, including robotics, AR/VR, and autonomous driving. To process the point cloud, many deep-learning-based point cloud recognition algorithms have been proposed. However, to meet the requirement of applications like autonomous driving, the algorithm must be fast enough, rendering accelerators necessary at the inference stage. But existing point cloud accelerators are still inefficient due to two challenges. First, the multi-layer perceptron (MLP) during feature computation is the performance bottleneck. Second, the feature vector fetching operation incurs heavy DRAM access. 

In this paper, we propose Pointer, an efficient Resistive Random Access Memory (ReRAM)-based point cloud recognition accelerator with inter- and intra-layer optimizations. It proposes three techniques for point cloud acceleration. First, Pointer adopts ReRAM-based architecture to significantly accelerate the MLP in feature computation. Second, to reduce DRAM access, Pointer proposes inter-layer coordination. It schedules the next layer to fetch the results of the previous layer as soon as they are available, which allows on-chip fetching thus reduces DRAM access. Third, Pointer proposes topology-aware intra-layer reordering, which improves the execution order for better data locality. Pointer proves to achieve 40$\times$ to 393$\times$ speedup and 22$\times$ to 163$\times$ energy efficiency over prior accelerators without any accuracy loss.

\end{abstract}


\title{\huge Pointer: An Energy-Efficient ReRAM-based \underline{Point} Cloud Recognition Accel\underline{er}ator with Inter-layer and Intra-layer Optimizations}

\author{ 
\large Qijun Zhang, Zhiyao Xie*\\ 
\large Hong Kong University of Science and Technology\\
\large qzhangcs@connect.ust.hk, eezhiyao@ust.hk
}

\begin{abstract}

Point cloud is an important data structure for a wide range of applications, including robotics, AR/VR, and autonomous driving. To process the point cloud, many deep-learning-based point cloud recognition algorithms have been proposed. However, to meet the requirement of applications like autonomous driving, the algorithm must be fast enough, rendering accelerators necessary at the inference stage. But existing point cloud accelerators are still inefficient due to two challenges. First, the multi-layer perceptron (MLP) during feature computation is the performance bottleneck. Second, the feature vector fetching operation incurs heavy DRAM access. 

In this paper, we propose Pointer, an efficient Resistive Random Access Memory (ReRAM)-based point cloud recognition accelerator with inter- and intra-layer optimizations. It proposes three techniques for point cloud acceleration. First, Pointer adopts ReRAM-based architecture to significantly accelerate the MLP in feature computation. Second, to reduce DRAM access, Pointer proposes inter-layer coordination. It schedules the next layer to fetch the results of the previous layer as soon as they are available, which allows on-chip fetching thus reduces DRAM access. Third, Pointer proposes topology-aware intra-layer reordering, which improves the execution order for better data locality. Pointer proves to achieve 40$\times$ to 393$\times$ speedup and 22$\times$ to 163$\times$ energy efficiency over prior accelerators without any accuracy loss.

\end{abstract}


\maketitle
\pagestyle{plain}
\begingroup\renewcommand\thefootnote{*}
\footnotetext{Corresponding Author}
\endgroup

\section{Introduction}

\begin{figure*}[!t]
\includegraphics[width=0.98\textwidth]{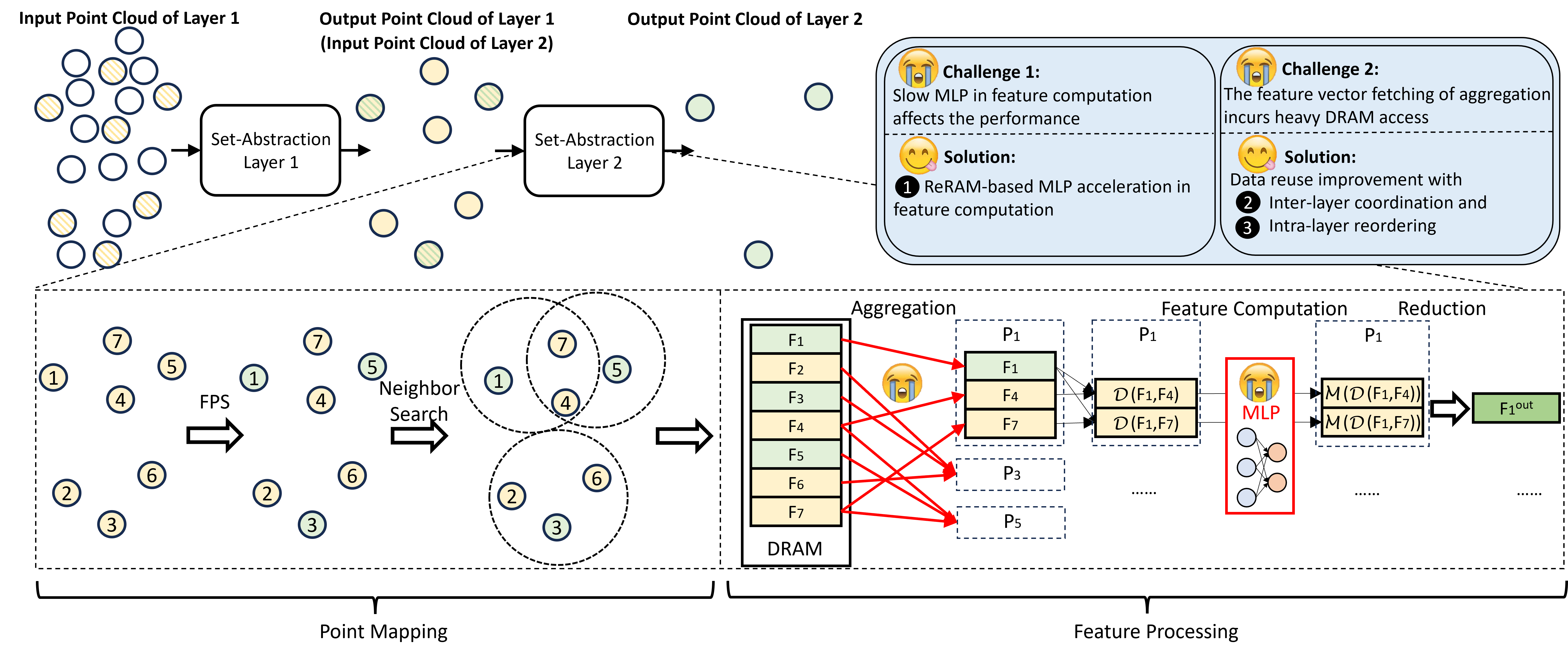}
\vspace{-.1in}
\caption{The workflow of PointNet++, which consists of two major stages named point mapping and feature processing.
The point mapping stage includes \emph{farthest point sample (FPS) and neighbor search}. The feature processing stage includes \emph{aggregation}, \emph{feature computation}, and \emph{reduction}. In the \emph{aggregation} step, for each sampled point $P_i$ with feature vector $F_i$, its neighboring points $P_j$'s feature vectors $F_j$ are also fetched. Then their difference $\mathcal{D}(F_i, F_j)$ is computed. Then an MLP $\mathcal{M}$ performs \emph{feature computation}, generating $\mathcal{M}(\mathcal{D}(F_i, F_j))$ for each $P_j$. Finally all results are reduced by computing the maximum of each column.}
\label{setabstractlayer}
\vspace{-.1in}
\end{figure*}

Point cloud is a data structure dedicated to describing three dimensional (3D) objects using coordinates and other auxiliary features~\cite{rusu20113d}. 
It is increasingly popular in many applications, including virtual reality, augmented reality, autonomous driving, and robotics. 
To process the point cloud data, multiple point cloud recognition algorithms have been proposed~\cite{qi2017pointnet, qi2017pointnet++}. PointNet++~\cite{qi2017pointnet++} and its variants are one of the most widely used point cloud recognition algorithms. As Fig.~\ref{setabstractlayer} shows, PointNet++ consists of multiple set-abstraction layers. Each set-abstraction layer includes two stages: 1) Point mapping stage, which performs the farthest point sample (FPS) and neighbor search to compute the mapping between output point cloud and input point cloud; 2) Feature processing stage, which first aggregates feature vectors according to the mapping, then computes features with multi-layer perceptron (MLP), finally reduces intermediate results to output feature vector.

However, most point-cloud applications require the point cloud recognition to be performed in real-time to enable interactions with humans or environments, bringing a long-lasting efficiency challenge in deployment. 
To solve this challenge, an energy-efficient high-performance point cloud accelerator is crucial. Existing accelerator designs include Mesorasi~\cite{feng2020mesorasi}, Crescent~\cite{feng2022crescent}, Point-X~\cite{zhang2021point}, PointAcc~\cite{lin2021pointacc}, PRADA~\cite{song2023prada}, FLNA~\cite{lyu2023flna}, TiPU~\cite{zheng2023tipu}, MARS~\cite{yang2023efficient}, Sava~\cite{liu2024sava}, and FusionArch~\cite{liu2024fusionarch}. However, despite these explorations~\cite{feng2020mesorasi, lin2021pointacc, zhang2021point, feng2022crescent, song2023prada, lyu2023flna, zheng2023tipu, yang2023efficient, liu2024sava, liu2024fusionarch}, as summarized in Fig.~\ref{setabstractlayer}, there are still two unsolved challenges in existing point cloud accelerators:
\begin{enumerate}
\item The first challenge is slow MLP execution in feature computation. MLP takes up about 70\% execution time when executed on GPUs~\cite{feng2020mesorasi}. Despite existing efforts in optimizing~\cite{song2023prada,feng2020mesorasi} or optimizing~\cite{lin2021pointacc} MLP executions, it is still the bottleneck. 
\item The second challenge is the heavy DRAM access in the aggregation step, as depicted in Fig~\ref{setabstractlayer}. For each point, it needs to fetch the feature vectors of all input points from DRAM. This operation incurs heavy DRAM access.
\end{enumerate}

To solve these two challenges, we propose Pointer, a ReRAM-based point cloud recognition accelerator with both inter- and intra-layer optimizations. As Fig~\ref{setabstractlayer} shows, it brings three essential improvements, denoted as \circled{1}\,\circled{2}\,\circled{3}. 
For challenge 1, Pointer proposes \circled{1} a ReRAM-based architecture to accelerate the MLP execution bottleneck~\cite{chi2016prime}\cite{shafiee2016isaac} in PointNet++. Such in-memory processing accelerates MLP by reducing costly data movement of weight fetching in this MLP operation. 
For challenge 2, Pointer reduces the heavy DRAM access by optimizing the dataflow across set-abstraction layers. 
Existing solutions first complete all computations in the previous layer and save results to DRAM, then fetch these results from DRAM as the inputs for the next layer. 
To reduce such DRAM access, Pointer proposes \circled{2} inter-layer coordination to enable on-chip fetching. It starts the calculation of the point in next layer as soon as all its required inputs from the previous layer are available. Such immediate data reuse allows on-chip storage and fetching of previous layer's results.
To further improve data locality, Pointer proposes \circled{3} topology-aware intra-layer reordering to reorder execution.  
The new execution order maximizes the reuse of the common inputs of processed points, further reducing DRAM accesses. 

The contributions in Pointer can be summarized as below. 
\begin{itemize}
    \item 
    To the best of our knowledge, Pointer is the first ReRAM-based accelerator for PointNet++-based point cloud recognition. ReRAM array greatly speeds up the slow and energy-hungry MLP execution during feature computation step. 
    \item 
    To reduce DRAM access, Pointer proposes inter-layer coordination.
    It schedules the next layer to fetch the results of the previous layer as soon as they are available, which allows on-chip fetching thus reduces DRAM access.
    \item 
    To further reduce DRAM access, Pointer proposes intra-layer reordering. It reschedules the execution order with topology awareness, further allowing data reuse of common inputs.
    \item
    Pointer achieves 40$\times$ to 393$\times$ speedup and 22$\times$ to 163$\times$ energy efficiency over the state-of-the-art no-accuracy-loss accelerator with a similar hardware cost.
\end{itemize}

\section{Background and Related Work}


\subsection{PointNet++: Deep Learning on Point Clouds}


PointNet++~\cite{qi2017pointnet++} is one of the most widely-adopted point cloud recognition algorithms. It consists of multiple set-abstraction layers. 
As Fig.~\ref{setabstractlayer} shows, each set-abstraction layer will transform an input point cloud to an output point cloud with fewer points, which belong to a subset of the original input. Then this output point cloud will be the input of the next set-abstraction layer. Each set-abstraction layer consists of two major stages, named point mapping and feature processing.

The point mapping stage consists of two steps, named \emph{farthest point sampling (FPS)} and \emph{neighbor search}. The FPS will determine which input points should remain in the output point cloud of this layer, as shown in Fig.~\ref{setabstractlayer}. As mentioned, for the output point cloud of each layer, its points are the subset of the input point cloud. 
The \emph{neighbor search} determines the neighbors of each output point by searching for the top-k nearest points of it.


The feature processing stage consists of three steps, named \emph{aggregation}, \emph{feature computation}, and \emph{reduction}. We use a point $P_i$ as an example, assuming it is selected to remain in the output cloud point. We denote its feature vector as $F_i$ and its neighboring points as $P_j$ with feature vector $F_j$. The \emph{aggregation} stage first calculates the ``difference" $\mathcal{D}(F_i,F_j)$ between feature vector of $P_i$ and each of its neighbors.  Then at the \emph{feature computation} stage, an multi-layer perceptron (MLP) $\mathcal{M}$ is applied on $\mathcal{D}(F_i,F_j)$ to generate $\mathcal{M}(\mathcal{D}(F_i,F_j))$. Then all of the $\mathcal{M}(\mathcal{D}(F_i,F_j))$ corresponding to all neighbors $P_j$ are reduced by maximum pooling, generating the output feature vector $F_{i}^{out}$ for point $P_i$.


\begin{figure}[!b]
\vspace{-.15in}
\includegraphics[width=0.35\textwidth]{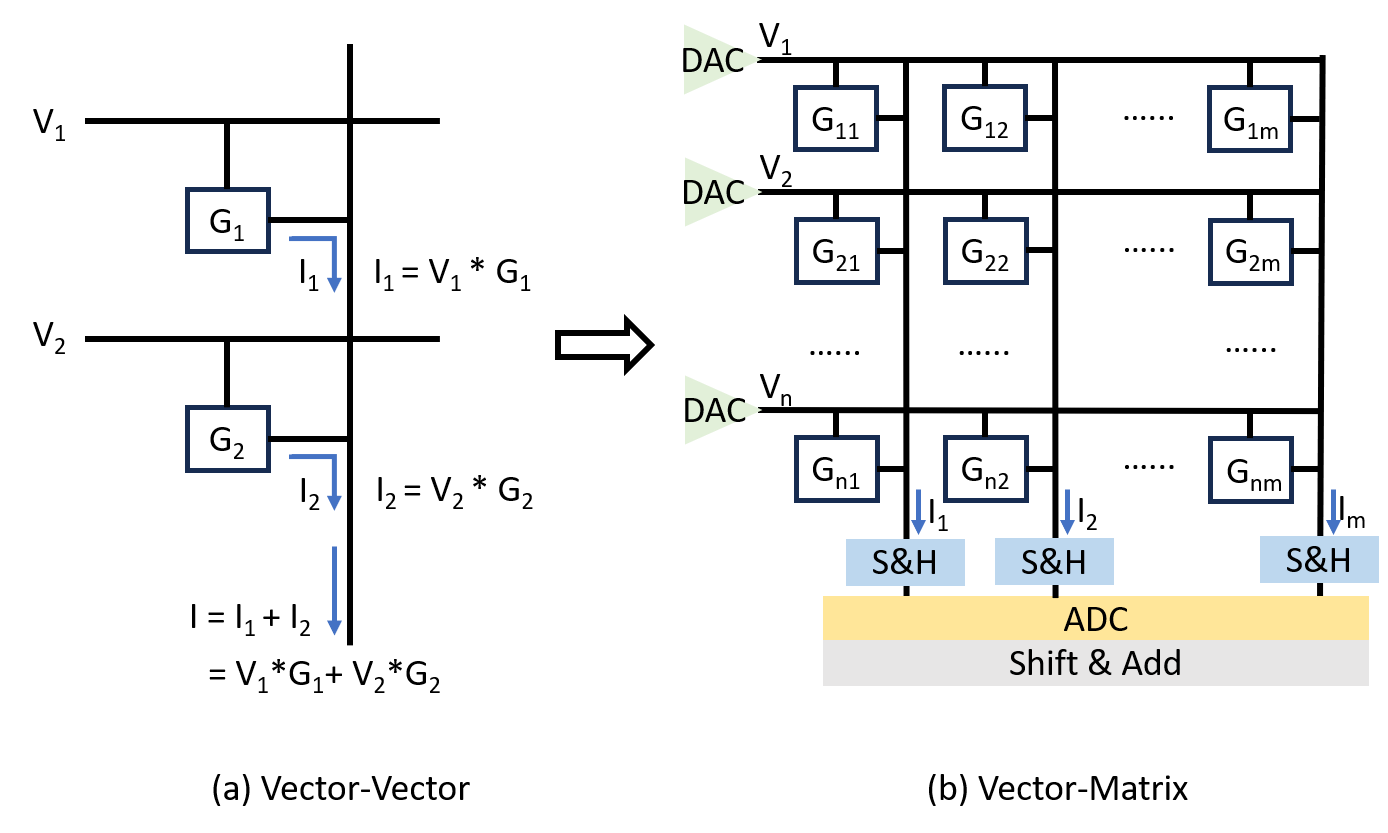}
\vspace{-.1in}
\caption{(a) Multiply-accumulate operation with ReRAM. (b) The ReRAM array used as vector-matrix multiplier.}
\label{reram}
\end{figure}

\subsection{Process-In-Memory with ReRAM}
\label{subsec:reram}


ReRAM is an emerging non-volatile memory technology. Besides storing data, the ReRAM crossbar architecture is widely adopted to accelerate the vector-matrix multiplication~\cite{chi2016prime,shafiee2016isaac}. The mechanism is shown in Fig.~\ref{reram}(a), with the vertical line denoting bitline, and horizontal line denoting wordline. The resistances of each memristor cell can be programmed for computation. Assuming the resistances of the two cells in Fig.~\ref{reram}(a) are programmed as $R_1$ and $R_2$, thus the conductance values are $G_1=1/R_1$ and $G_2=1/R_2$. When the voltages of these wordlines are $V_1$ and $V_2$, thus the current on the bitline is $I=I_1+I_2=V_1*G_1+V_2*G_2$. Therefore, if we set conductance values $\{G_1,G_2, ..., G_n\}$ equal to $n$ elements in a vector and voltages $\{V_1,V_2, ..., V_n\}$ equal to the other vector, then the current on the bitline $I$ is the dot product result of two vectors.


To extend the vector product engine to a vector-matrix multiplier, which is the fundamental operation in MLP, the structure in Fig.~\ref{reram}(a) can be horizontally extended to a crossbar architecture in Fig.~\ref{reram}(b). Assume the vector length is still $n$ and the matrix shape is $n\times m$. 
In this case, the $m$ current values on the $m$ bitlines $\{I_1,I_2,...,I_m\}$ can represent the output vector.

\subsection{Related Work}

Existing accelerators for PointNet++-based point cloud recognition algorithms can be categorized into two types, depending on whether they incur accuracy variation. Accelerators with accuracy variation include Mesorasi~\cite{feng2020mesorasi}, Crescent~\cite{feng2022crescent}, PRADA~\cite{song2023prada}, FLNA~\cite{lyu2023flna}, TiPU~\cite{zheng2023tipu}, Sava~\cite{liu2024sava}, and FusionArch~\cite{liu2024fusionarch}. But accuracy variation is often unacceptable, especially in accuracy-critical scenarios where the accuracy variation can lead to catastrophic consequences. The accelerators without accuracy variation include Point-X~\cite{zhang2021point}, PointAcc~\cite{lin2021pointacc}, MARS~\cite{yang2023efficient}. The Point-X only focuses on DGCNN~\cite{wang2019dynamic}, which cannot be directly applied to widely-adopted PointNet++ and its other variants. In other words, it does not generalize well. Only PointAcc and MARS can accelerate general point cloud recognition algorithms without accuracy variation, and MARS mainly enhances the mapping unit of PointAcc. Our Pointer is also a general point cloud accelerator without accuracy variation. So in this paper, the baseline is MARS-like accelerator, which is a state-of-the-art no-accuracy-variation point cloud accelerator. 

Although HSPA~\cite{ling2024rram} proposed a ReRAM-based point cloud recognition accelerator, it focuses on the DeepSets~\cite{zaheer2017deep} algorithms. It can not fully support the PointNet++-based algorithms, where irregular DRAM access is critical even with the ReRAM engine. Moreover, although Point-X~\cite{zhang2021point}, TiPU~\cite{zheng2023tipu}, and FusionArch~\cite{liu2024fusionarch} discussed the ``spatial-locality'' and packed adjacent point processing together spatially or temporally, they did not consider the global schedule space where the inter-layer coordination is applied, which is hard to support directly by their architecture design. In comparison, a simple reordering with light hardware overhead proposed in our work can be natural when integrating with the inter-layer coordination which is also an order-related technique. These two techniques can be achieved in a scheduler uniformly.

\section{Methodology}


\begin{figure*}[!t]
\includegraphics[width=0.98\textwidth]{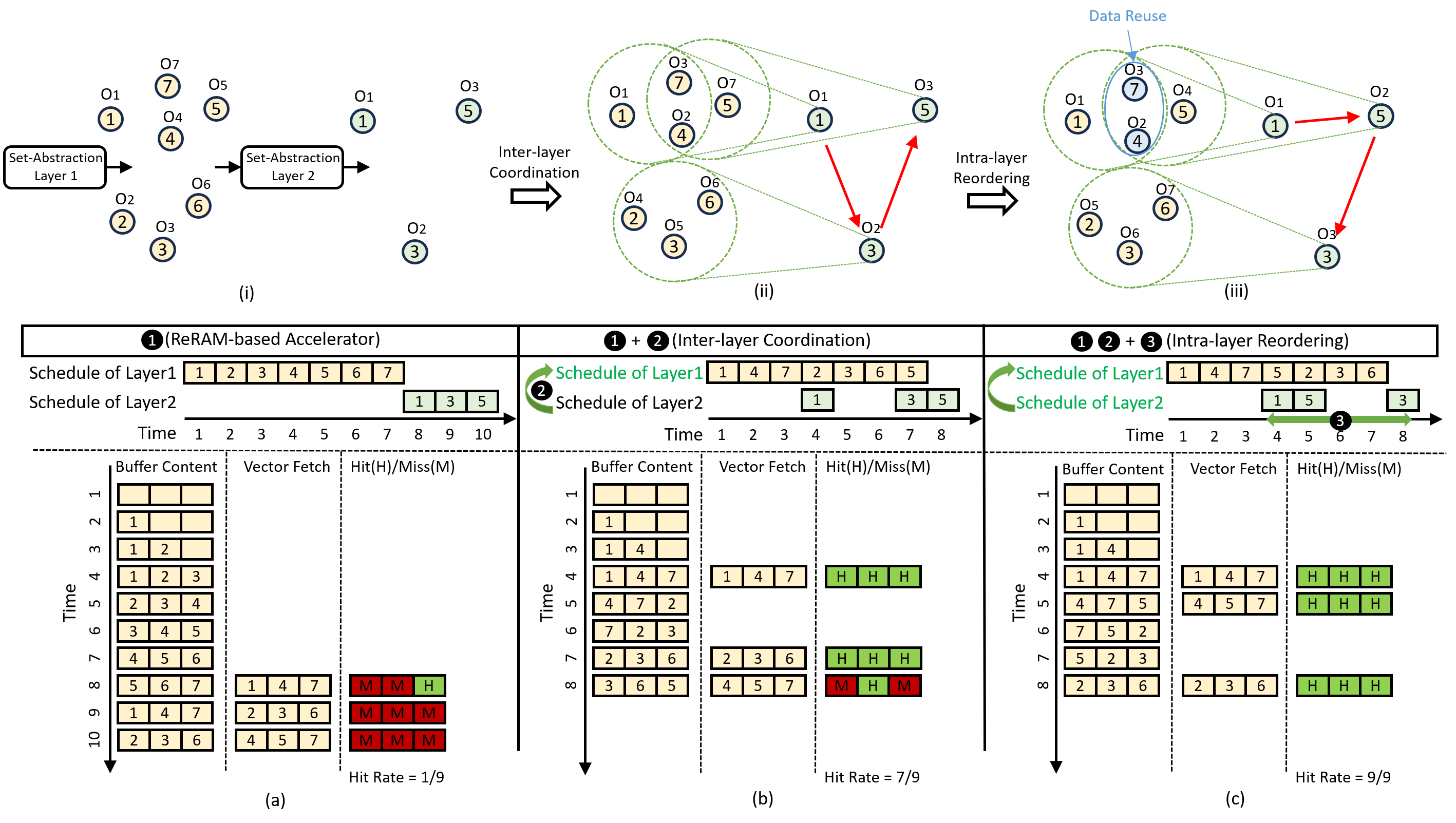}
\vspace{-.15in}
\caption{Inter-layer coordination and intra-layer reordering, using the example in Fig.~\ref{setabstractlayer}. The upper sub-figures (i)(ii)(iii) illustrate the point execution order in each layer. The number within circles is the index of points $P_i$, and the $O_i$ outside the circles is the process order within each layer. The bottom sub-figures (a)(b)(c) illustrate how inter-layer coordination and intra-layer reordering improve the on-chip buffer hit rate thus performance. The buffer content shows the available content at the start point of each time step. (a) Basic ReRAM-based acceleartor\protect\footnotemark. It simply schedules the execution by index order. (b) Accelerator with inter-layer coordination. It schedules the execution order in layer 1 based on the receptive field of points in layer 2. (c) Accelerator with both inter-layer coordination and intra-layer reordering (i.e., Pointer). The intra-layer reordering determines the execution order of layer 2. The inter-layer coordination still determines the execution order of layer 1.} 
\label{workflow}
\vspace{-.1in}
\end{figure*}



In this Section, we will present Pointer in detail. We first cover the basic ReRAM-based PointNet++ accelerator design, which greatly accelerates the MLP operation. Then we introduce our proposed inter-layer coordination and topology-aware intra-layer reordering techniques, which further effectively reduce the DRAM access. Finally, we describe the detailed hardware implementation of Pointer.

\subsection{Basic ReRAM-based Point Cloud Accelerator}
\label{basic}

Conventional PointNet++ accelerators commonly use Multiply-and-Accumulate (MAC) array to compute MLP. Because of the limited on-chip buffer, it requires repeatedly loading the weight from DRAM, leading to the slow and energy-hungry execution of MLP. In comparison, ReRAM is a promising solution for vector-matrix multiplication. By performing in-memory computing, the data movement overhead in existing MLP is eliminated, which improves performance and reduces energy consumption. This leads to the basic ReRAM-based PointNet++ accelerator. Similar to the existing dataflow in Fig.~\ref{setabstractlayer}, the feature vectors are fetched and differences are computed, then fed into ReRAM-based MLP engine, finally the feature vector of output point is written into the DRAM.

\footnotetext{Here we assume there is a simple buffer in the basic ReRAM-based accelerator, in order to compare with designs with inter-layer coordination and intra-layer reordering.}

Reliability is a main concern when adopting ReRAM-based accelerators. For better reliability, we propose to adopt the ReRAM array with a relatively small number of bits per cell (i.e. 2 bits per cell). In addition, since the ReRAM computation is not the speed bottleneck, we can trade off the ReRAM computation speed for lower overhead. For example, we can accept more sequential operations by adopting fewer ReRAM array replications~\cite{shafiee2016isaac}.


Because different set-abstraction layers are mapped to different ReRAM arrays, different layers can be theoretically executed in parallel. But in fact, because of complex data dependency between layers, for basic ReRAM-based design, each set-abstraction layer is still executed sequentially, as shown in Fig.~\ref{workflow}(a). In other words, the next set-abstraction layer still starts after all computation in the previous layer is completed. And for each set-abstraction layer, following existing solution, output points are still calculated directly by their index orders. For example, if output points are $\{P_1,P_3,P_5,P_7,P_9\}$, then at this layer, execution order is $O = [P_1-P_3-P_5-P_7-P_9]$. This naive scheduling will be improved in subsequent subsections. 


\subsection{Enable On-Chip Fetching with Inter-layer Coordination}



After adopting ReRAM, the MLP execution is no longer the runtime bottleneck. The new bottleneck becomes the heavy DRAM access during feature vector fetching in aggregation, which also incurs high energy consumption for DRAM.

DRAM access of feature vector fetching can be reduced by buffering the data and fetching them on-chip, which can not only improve performance but also reduce energy consumed by DRAM. However, since the existing design requires previous set-abstraction layer to be fully completed before starting the next layer, it will require all feature vectors to be held on-chip till the fetching, requiring a prohibitively large buffer. Otherwise, feature vectors generated by previous layer will be evicted into DRAM before they are requested by the next layer. We illustrate this scenario in Fig.~\ref{workflow}(a).

\newlength{\textfloatsepsave} \setlength{\textfloatsepsave}{\textfloatsep} \setlength{\textfloatsep}{0pt} 
\begin{algorithm}[!b]\small
\caption{Scheduling Order Generation}
\label{alg3}
\begin{flushleft}
  \textbf{Input}: The number of layers, $l$. All output points of the last layer, $SP$. The distance function between two points, $T(P_i,P_j)$. Receptive fields of each point $\mathbf{E_i^k}-\{E^{k-1}_j\}$\\
  \textbf{Output}: The execution order of each layer $\{O_1,O_2,...O_l\}$
\end{flushleft}

\begin{algorithmic}[1]
\Statex \emph{/* line 1-8 are algorithm of \circled{3} Intra-layer reordering */}
\Statex \emph{/* Select points one by one based on the topology */}
\State Initiate an empty order list for the last layer $O_l = []$ \label{line1}
\Statex \emph{/* Start from a random point */}
\State Select a point $P_i$ from $SP$ randomly
\State Take $P_i$ out from $SP$, append it to $O_l$
\State Set $LP$ = $P_i$
\While{\ $SP$ is not empty}
    \Statex \emph{\ \ \ \ \ /* Select point nearest to the last selected point*/}
    \State $CP = \mathop{\arg\min}\limits_{P_j \in SP} T(LP,P_j)$
    \Statex \emph{\ \ \ \ \ /* $CP$ denotes the current selected point*/}
    \Statex \emph{\ \ \ \ \ /* $LP$ denotes the last selected point*/}
    \State Take $CP$ out from $SP$, append it to $O_l$
    \State Set $LP$ = $CP$
\EndWhile \label{line2}

\Statex 

\Statex \emph{/* line 9-13 are algorithm of \circled{2} Inter-layer coordination */}
\Statex \emph{/* The execution order of prior layers depend on later layers, thus it iterates from later layer to prior layers */}
\For {$k:\ from\ l-1\ to\ 1\ (descend)$} \label{line3}
    \Statex \emph{\ \ \ \ \ /* Generate the order for layer $k$*/}
    \State Initiate an empty order list for layer $k$, $O_k = []$
    \For {$E_j^{k+1} \in O_{k+1}$ (by order)}
    \State Read its receptive field $\mathbf{E_j^{k+1}}-\{E^{k}_m\}$
    \State $O_{k}$.append($\{E^{k}_m\}$) 
    \EndFor
\EndFor \label{line4}
\end{algorithmic} 
\end{algorithm}
\setlength{\textfloatsep}{\textfloatsepsave}

To address this DRAM access bottleneck, we propose the inter-layer coordination technique to improve on-chip fetching. The key idea is to start the computation of the point in the next layer earlier, immediately after the calculations of all its required input points from the previous layer have finished. It requires coordinating the execution of the previous layer based on the execution order of the next layer. For such optimization, the dependency between the points in different layers is critical.

\begin{figure}[!t]
\centering
\includegraphics[width=0.48\textwidth]
{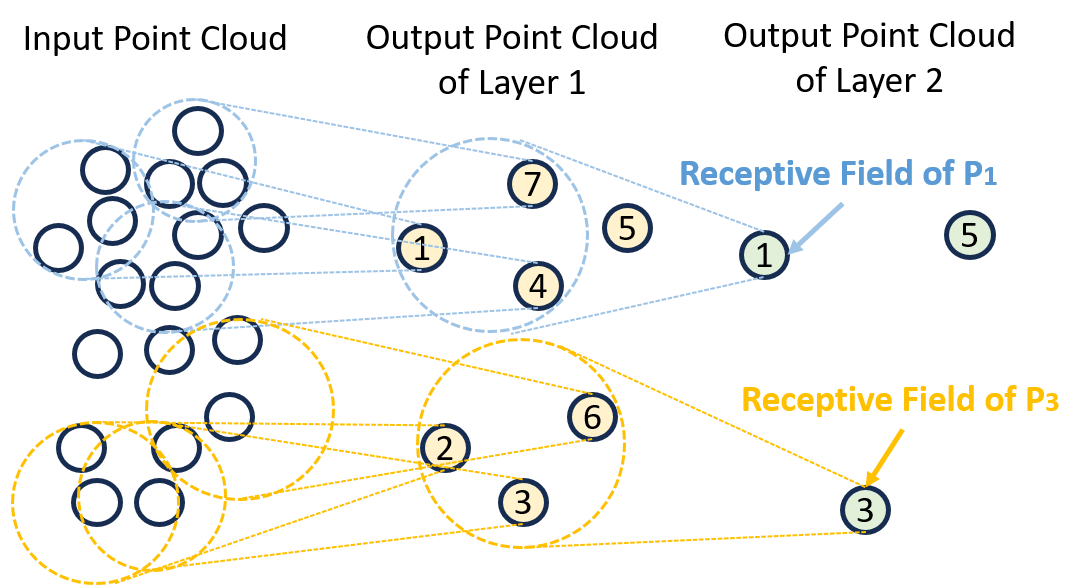}
\caption{An example of the pyramid-shaped receptive field. This example is consistent with Fig.~\ref{setabstractlayer} and Fig.~\ref{workflow}.} 
\label{slidingwindow}
\vspace{-.2in}
\end{figure}

As Fig.~\ref{slidingwindow} shows, the dependency across multiple consecutive layers leads to a pyramid-shaped receptive field for each last-layer output point, which can be viewed as the top of the ``pyramid''. In this example, two set-abstraction layers will lead to a 3-level pyramid-shaped receptive field.

The algorithm of inter-layer coordination is shown as the lines~\ref{line3}-\ref{line4} of Algorithm~\ref{alg3}. The example of inter-layer coordination is illustrated in Fig.~\ref{workflow}(ii)(b) where the number of layer $l$ is 2. We denote the computation of point $P_i$ in the layer $j$ as $E_{i}^{j}$. For the example with two set-abstraction layers, we represent a pyramid-shaped receptive field as $\mathbf{E_i^2}-\{E^1_j\}$, where $\mathbf{E_i^2}$ represents the executions of $P_i$ in layer 2 and $\{E^1_j\}$ represents the execution of all points in $P_i$'s receptive field in layer 1. To distinguish two different layers in our example, we use bold type to emphasize executions of points in layer 2. In our example, corresponding to the three output points, there are three pyramid-shaped receptive fields: (1)$\mathbf{E_1^2}-\{E_1^1,E_4^1,E_7^1\}$, (2)$\mathbf{E_3^2}-\{E_2^1,E_3^1,E_6^1\}$, (3)$\mathbf{E_5^2}-\{E_4^1,E_5^1,E_7^1\}$. Such receptive field information determines the data dependency.

We schedule the computation to maximize data reuse according to the data dependency. As the example shown in Fig.~\ref{workflow}(ii)(b), the computation is now receptive-field by receptive-field, instead of layer by layer. Because the computation order of the last layer (layer 2) is the index order by default, which means that $O_2 = [P_1-P_3-P_5]$, equivalent to $O_2 = [\mathbf{E_1^2}-\mathbf{E_3^2}-\mathbf{E_5^2}]$. So such receptive-field by receptive-field scheduling first calculates 
$E_1^1-E_4^1-E_7^1-\mathbf{E_1^2}$, then $E_2^1-E_3^1-E_6^1-\mathbf{E_3^2}$, then $E_4^1-E_7^1-E_5^1-\mathbf{E_5^2}$. But actually $E_4^1$ and $E_7^1$ appear in two receptive fields and they only need to be calculated once. When we schedule these three receptive fields directly using the index order in the second layer $\mathbf{E_1^2}-\mathbf{E_3^2}-\mathbf{E_5^2}$, the final execution order with both layers is shown below: 
\begin{equation}
E_1^1-E_4^1-E_7^1-\mathbf{E_1^2}-E_2^1-E_3^1-E_6^1-\mathbf{E_3^2}-E_5^1-\mathbf{E_5^2} \label{eq:interlayer}    
\end{equation}
It can also be expressed as $O_1 = [E_1^1-E_4^1-E_7^1-E_2^1-E_3^1-E_6^1-E_5^1], O_2 = [\mathbf{E_1^2}-\mathbf{E_3^2}-\mathbf{E_5^2}]$. Since the executions in layer 1 and layer 2 can be executed in parallel in the ReRAM-based accelerator, this final order in Equation~\ref{eq:interlayer} is the same as the schedule of both layers in Fig.~\ref{workflow}(b).

In summary, the inter-layer coordination reduces DRAM access by holding feature vectors on-chip for immediate subsequent fetching. 
But as shown in Fig.~\ref{workflow}(b), there are still some on-chip buffer misses when the $E_4^1$ and $E_7^1$ are fetched at the second time.
It indicates that inter-layer coordination alone is still insufficient.


\subsection{Topology-aware Intra-layer Reordering}

The limitation of only using inter-layer coordination is that current scheduling is unaware of point cloud topology. 
It simply schedules the execution of the whole receptive field based on the index order of points in the last layer (i.e., $O_2 = [\mathbf{E_1^2}-\mathbf{E_3^2}-\mathbf{E_5^2}]$), without considering the actual topology of the point cloud. As a result, consecutive executed receptive fields have small overlap thus poor data locality. Using the example in Fig.~\ref{workflow}(ii), 
there is no overlap between receptive fields of $\mathbf{E_1^2}$ and $\mathbf{E_3^2}$, and also that of $\mathbf{E_3^2}$ and $\mathbf{E_5^2}$. 


To tackle the problem, we propose topology-aware intra-layer reordering to schedule the execution order of the last layer (i.e., layer 2 in this example). The topology-aware intra-layer reordering is shown as the lines~\ref{line1}-\ref{line2} of Algorithm~\ref{alg3}. The topology-aware order means that rather than scheduling them by index order, we generate a new order that tries to schedule consecutive points to be neighboring in the physical space. In this example, it reorders the scheduling of points in the last layer from the original $O_2 = [\mathbf{E_1^2}-\mathbf{E_3^2}-\mathbf{E_5^2}]$ to the $O_2' = [\mathbf{E_1^2}-\mathbf{E_5^2}-\mathbf{E_3^2}]$. 
Since inter-layer coordination makes a receptive-field by receptive-field scheduling, executions in previous layers in the receptive field will follow the order in the last layer. The new overall order will now become:
\begin{equation}
E_1^1-E_4^1-E_7^1-\mathbf{E_1^2}-E_5^1-\mathbf{E_5^2}-E_2^1-E_3^1-E_6^1-\mathbf{E_3^2} \label{eq:intralayer}    
\end{equation}
as illustrated in Fig.~\ref{workflow}(c). It can also be expressed as $O_1 = [E_1^1-E_4^1-E_7^1-E_5^1-E_2^1-E_3^1-E_6^1], O_2 = [\mathbf{E_1^2}-\mathbf{E_5^2}-\mathbf{E_3^2}]$. It now successfully removes all on-chip buffer misses by further exploiting data locality and boosting data reuse. The improved data reuse can further reduce DRAM access, thus improving performance and energy efficiency.

However, a challenge in this intra-layer reordering is how to find out which points in the last layer have the largest common part in their receptive fields. The most intuitive solution is to search all of the possible orders and analyze the data reuse of each order, which is unrealistic. The accelerator should explore an efficient way to generate the execution order.




\begin{figure}[!t]
\includegraphics[width=0.49\textwidth]{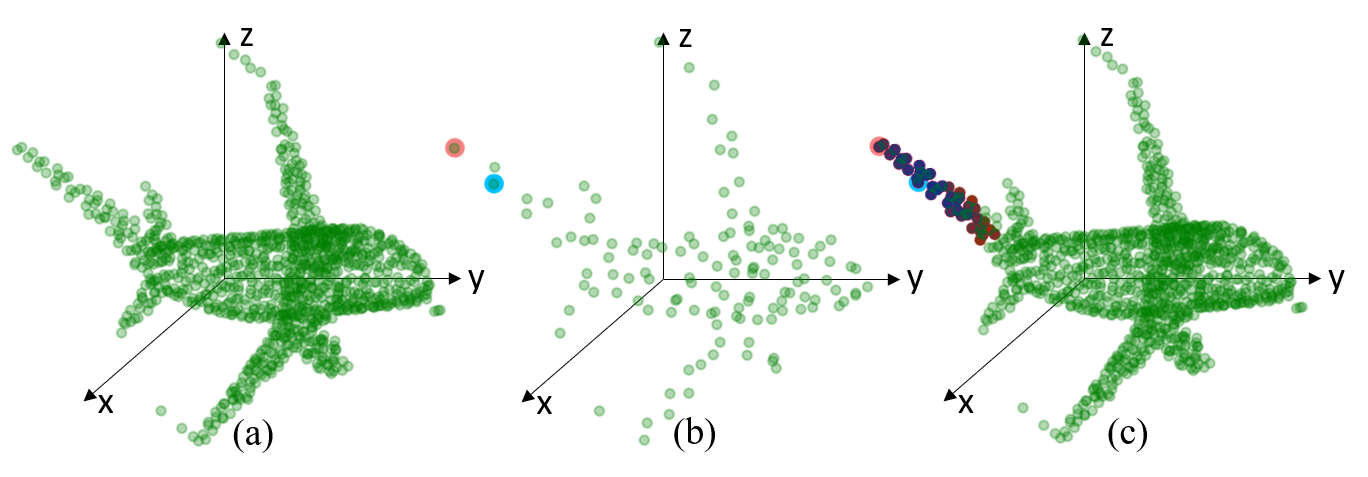}
\vspace{-.1in}
\caption{An obvious overlap between the receptive fields of two neighboring points in the last layer. (a) Original point cloud. (b) Green points are the output points in the last set-abstraction layer, the red and blue points are two neighboring points. (c) Green points are original point cloud, the red and blue points are the receptive fields of the two points in (b) respectively. There is a large overlap between these two fields. 
}
\label{overlap}
\end{figure}

\begin{figure}[!t]
\includegraphics[width=0.49\textwidth]{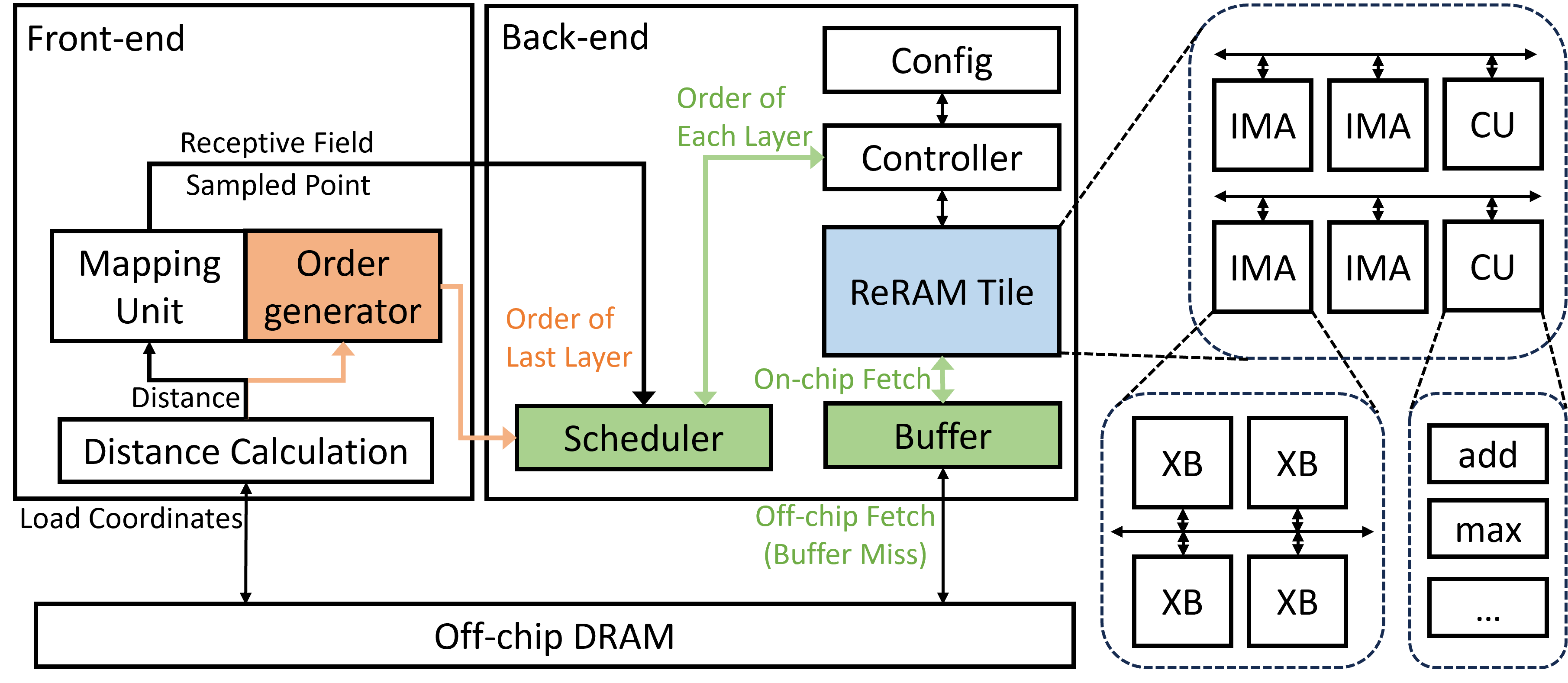}
\vspace{-.1in}
\caption{The architecture of Pointer. The blue part is the support for the ReRAM-based accelerator, the green part is for inter-layer coordination, and the orange part is for intra-layer reordering.}
\label{pointrdesign}
\vspace{-.1in}
\end{figure}

In Pointer, we adopt a highly lightweight yet effective technique to generate topology-aware execution order of points in the last layer. For all unexecuted points in the last layer, the (i+1)$^\text{th}$ point in the order should be the nearest to the i$^\text{th}$ point. Since these points are close in the last layer, points in their receptive field in previous layers are expected to also be close and overlap well. This is validated by an example from our experiment in Fig.~\ref{overlap}, which randomly selects two consecutive points in our generated topology-aware order, as shown in Fig.~\ref{overlap}(b). We find that their receptive fields in blue and red overlap very well, as Fig.~\ref{overlap}(c) shows. It indicates that this approximation can still generate a high-quality execution order. In addition, this reordering technique introduces negligible overhead, since it only requires the distance between selected pair of points, which fortunately has already been calculated during the existing FPS and neighbor search steps. 




\subsection{Hardware Implementation of Pointer}


Fig.~\ref{pointrdesign} shows the overview of the Pointer architecture design. The basic ReRAM-based PointNet++ accelerator introduced in Section~\ref{basic} is the part without color. It consists of front-end and back-end. Similar to the prior work~\cite{song2023prada}, the front-end is for the point mapping stage, and the back-end is for the feature processing stage. The back-end mainly consists of four parts: ReRAM tile, reconfigurable data path, digital computation unit, and main controller. 1) The ReRAM tile consists of many ReRAM arrays, which serve the computation and storage for weight in MLP. 2) The reconfigurable data path will control and transfer data between different ReRAM arrays.
3) The digital computation unit executes some operations such as ADD, MAX, and non-linear function. 4) The main controller coordinates different parts of the accelerator. 

The architecture support for inter-layer coordination is shown as the green part in Fig.~\ref{pointrdesign}. 
The support for topology-aware intra-layer reordering is shown as the orange part of Fig.~\ref{pointrdesign}. It is a small order generator added to front-end with negligible hardware overhead.

\section{Experiment}

\subsection{Experiment Setup}

\subsubsection{Benchmarks}
We evaluate the representative point cloud recognition model PointNet++~\cite{qi2017pointnet++} on our accelerator. Three different configurations of evaluated PointNet++ are summarized in Table.~\ref{modelconfig}. Same as the original PointNet++, all three models consist of two set-abstraction layers. 
The input point cloud size is 1024 points in our models. For dataset, we adopt ModelNet40~\cite{wu20153d}, which is widely adopted in many point cloud recognition research and point cloud accelerator evaluations, consisting of 12311 point clouds.

\begin{table}[!t]
      \centering
      \renewcommand{\arraystretch}{1.1}
      \resizebox{0.48\textwidth}{!}{
        \begin{tabular}{ |c|c||c|c|c| } 
        \hline
        & Model ID & Model 0 & Model 1 & Model 2 \\
        \hline
        \hline
        \multirow{7}{*}{Layer 1} & Input Feature Vector Length & 4 & 8 & 16 \\
        \cline{2-5}
        & Output Feature Vector Length & 128 & 256 & 512 \\
        \cline{2-5}
        & \multirow{3}{*}{The Shape of MLP (three layers)} & 4*64 & 8*128 & 16*256 \\
         & & 64*64 & 128*128 & 256*256 \\
         & & 64*128 & 128*256 & 256*512 \\
        \cline{2-5}
        & The Number of Neighbors & 16 & 16 & 16 \\
        \cline{2-5}
        & The Number of Central Point\footnotemark{} & 512 & 512 & 512 \\
        \hline
        \multirow{7}{*}{Layer 2} & Input Feature Vector Length & 129 & 256 & 512 \\
        \cline{2-5}
        & Output Feature Vector Length & 256 & 512 & 1024 \\
        \cline{2-5}
        & \multirow{3}{*}{The Shape of MLP (three layers)} & 128*128 & 256*256 & 512*512 \\
         & & 128*128 & 256*256 & 512*512 \\
         & & 128*256 & 256*512 & 512*1024 \\
        \cline{2-5}
        & The Number of Neighbors & 16 & 16 & 16 \\
        \cline{2-5}
        & The Number of Central Point & 128 & 128 & 128 \\
        \hline
        \end{tabular}
        }
        \caption{Three PointNet++ models evaluated in experiment.}
        \label{modelconfig}
        \vspace{-.3in}
\end{table}

\footnotetext{The number of central points is the number of selected output points in FPS.}

\subsubsection{Modeling Accelerator Architecture}
To evaluate our design, we develop a simulator to model the behavior of our design assuming 8GB/s DDR3 bandwidth. 
Specifically, we mainly simulate the back-end (i.e., feature processing stage) of the Pointer and our baseline. It is because when deployed in the application, the point mapping and feature processing stages can be pipelined and the feature processing is slower than point mapping.

Our design is evaluated under 40nm technology and the frequency is set to 1GHz. As for area, we use CACTI~\cite{muralimanohar2009cacti} to model the SRAM and published data from the prior work~\cite{shafiee2016isaac} to model ReRAM. The evaluated area of the back-end and the order generator of our design\footnote{The front-end of our design is similar to the baseline except for the order generator.} is $1.25mm^2$. For our design configuration, the ReRAM tile consists of 96 IMAs with each IMA consisting of 8 128*128 ReRAM arrays, and the buffer size is 9KB. We estimate energy efficiency with reference energy data collected from \cite{shafiee2016isaac, muralimanohar2009cacti}.


We select a baseline accelerator similar to MARS~\cite{yang2023efficient}. The baseline's MAC array consists of 32*32 MAC. For a fair comparison, we always keep the SRAM size used in our design and baseline the same, which is a 9KB on-chip buffer. The area of the back-end of the MARS-like baseline accelerator is $1.56mm^2$, which indicates that the hardware overhead of our design is similar to the baseline.

To further clearly evaluate each of our proposed methods, we also introduce two variants of Pointer for ablation study. The first one is the Pointer without inter-layer coordination and intra-layer reordering, which is called Pointer-1 (i.e., only with contribution \circled{1}). The second one is the Pointer without intra-layer reordering, which is called Pointer-12 (i.e., only with contribution \circled{1}\,\circled{2}).

\subsection{Experimental Result}

\subsubsection{Performance Speedup and Energy Efficiency}

\begin{figure}[!t]
\centering
\includegraphics[width=0.37\textwidth]{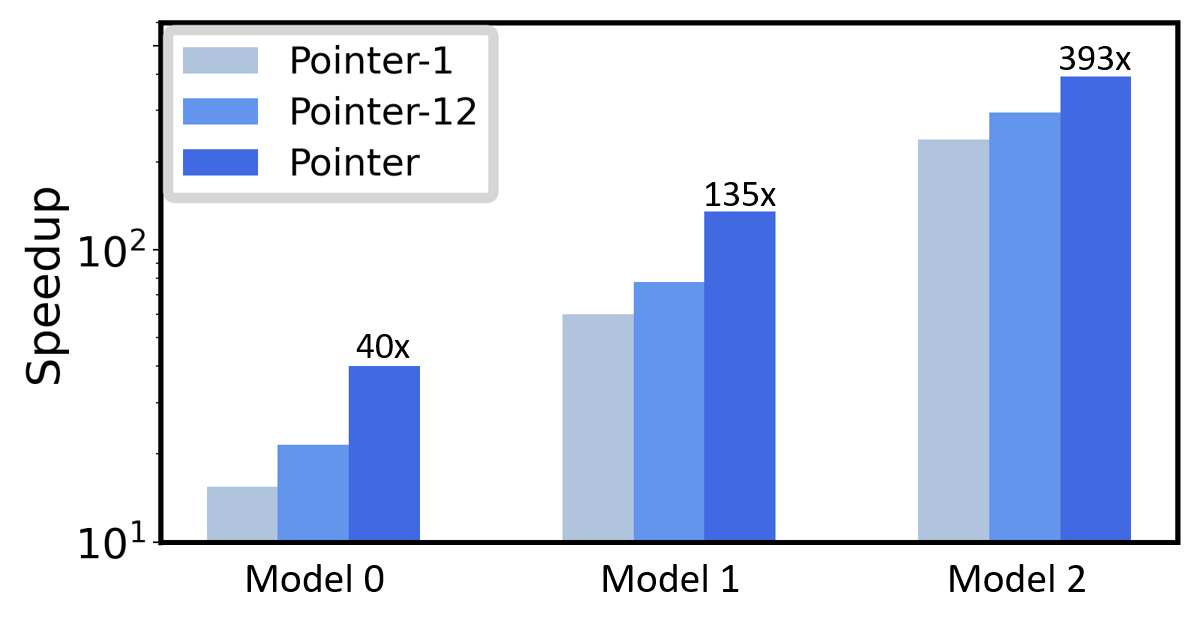}
\vspace{-.13in}
\caption{The speedup of Pointer and its two variants (i.e., Pointer-1, Pointer-12), compared with the MARS-like baseline~\cite{yang2023efficient}. It measures the performance of three different Point++ models, whose configurations are shown in Table~\ref{modelconfig}.}
\label{perforamce}
\vspace{-.13in}
\end{figure}

Fig.~\ref{perforamce} shows the speedup of Pointer over the MARS-like baseline. 
For three evaluated PointNet++ models, Pointer speedups by 40$\times$, 135$\times$, and 393$\times$. 
This speedup is more obvious for larger models, demonstrating the great scalability of Pointer. 
When the model scales up, the weight matrix becomes larger, so the data movement overhead in the baseline is heavier. But for Pointer, because the data movement overhead has been eliminated by using ReRAM, the performance slowdown is much slower than baseline, so the speedup is becoming significant.

The result also shows that the performance of Pointer-12 always outperforms Pointer-1. It validates the benefit of inter-layer coordination. The Pointer also always outperforms the Pointer-12. It validates the benefit of intra-layer topology-aware reordering. 

\begin{figure}[!t]
\centering
\includegraphics[width=0.37\textwidth]{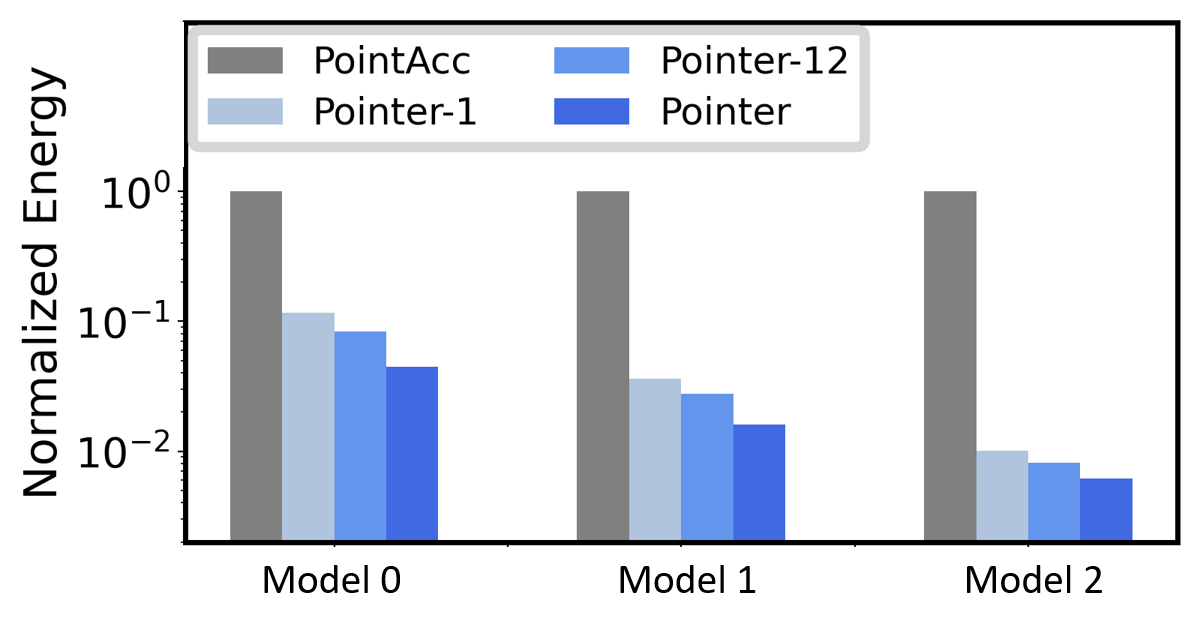}
\vspace{-.13in}
\caption{The energy consumption of Pointer and its two variants normalized with MARS-like baseline.}
\label{energy}
\vspace{-.13in}
\end{figure}

Fig.~\ref{energy} shows energy consumption of Pointer compared with the MARS-like baseline. The energy consumption is normalized with the baseline. For the three evaluated models, Pointer improves the energy efficiency by 22$\times$, 62$\times$, and 163$\times$, which demonstrates the energy efficiency improvement of Pointer. Because the energy consumption mainly comes from the DRAM access, the reduction of DRAM access in Pointer can significantly reduce the energy.

\begin{figure}[!b]
\vspace{-.15in}
\subfigure[]{
    \centering
    \includegraphics[width=0.46\columnwidth]{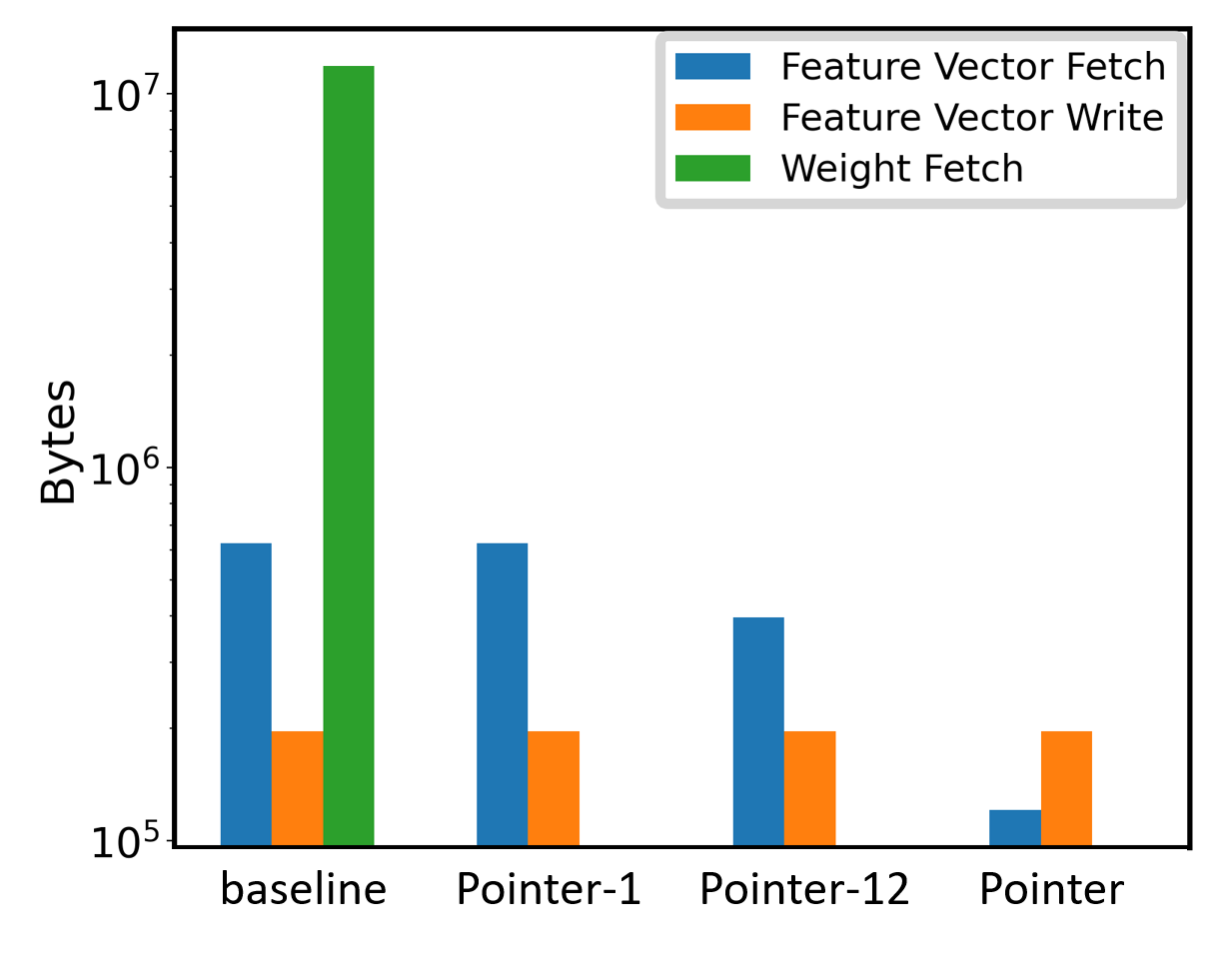}
    \label{breakdown}
}
\subfigure[]{
    \centering
    \includegraphics[width=0.46\columnwidth]{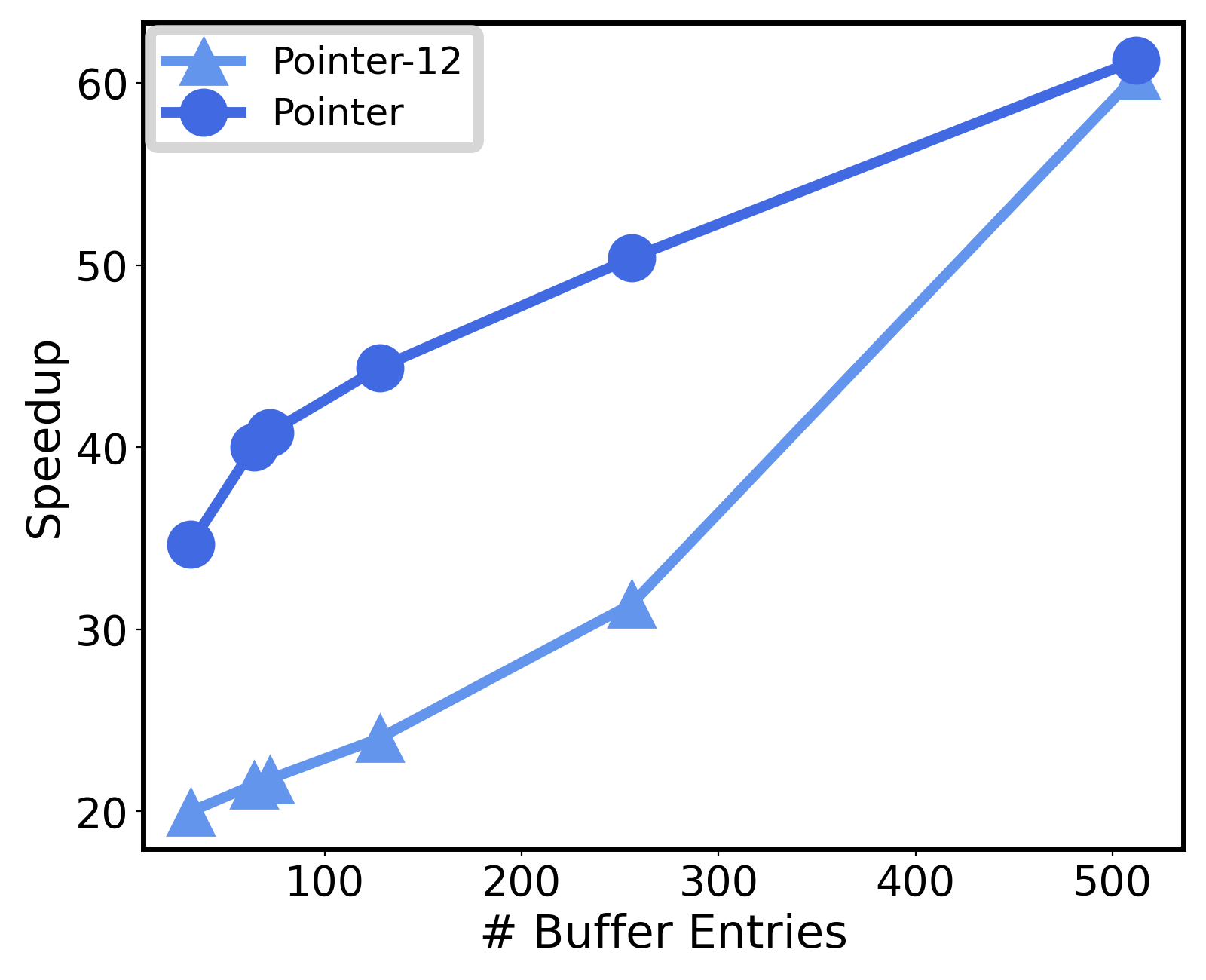}
    \label{perfsize}
}
\vspace{-.13in}
\caption{(a) The breakdown of DRAM traffic for feature vector fetching, feature vector writing, and weight fetching of Pointer, baseline, and two variants of Pointer for ablation study. (b) The comparison of Pointer-12 and Pointer for how the performance changes with the buffer size. There is no buffer for Pointer-1 so it is not shown.}
\label{resultanalysis}
\end{figure}


\subsubsection{Source of Performance Gain}
\label{subsec:cachehit}
To further analyze Pointer's superior performance, we present a breakdown of the overall DRAM access traffic into three parts: feature vector fetching, feature vector writing, and weight fetching for MLP, as shown in Fig.~\ref{breakdown}. 


In Fig.~\ref{breakdown}, the comparison between Pointer-1 and baseline shows that the ReRAM-based accelerator eliminates weight fetching. Compared with Pointer-1, the average DRAM traffic of feature vector fetching in Pointer-12 is reduced from 627KB to 396KB, cutting down 37\% of traffic by enabling on-chip fetching. Please notice that the feature vector writing remains unchanged, because all of the computed feature vectors will be saved back into the DRAM once. Compared with Pointer-12, the average DRAM traffic of feature vector fetching in Pointer is further reduced to 121KB, further cutting down 69\% of the traffic, also 81\% compared with Pointer-1, because the intra-layer reordering improves the data reuse. 

To further inspect how the intra-layer reordering reduces the feature vector fetching, we analyze the on-chip buffer hit rate with and without intra-layer reordering to demonstrate the data locality improvement. It shows that after equipping the accelerator with intra-layer reordering, the on-chip buffer hit rate of set-abstraction layer 1 is improved from 68\% to 71\%, and that of layer 2 is improved from 33\% to 82\%, which shows that data locality is improved.

\subsubsection{Impact of Buffer Size}

\begin{figure}[!t]
\vspace{-.15in}
\subfigure[Hit rate in layer 1]{
    \centering
    \includegraphics[width=0.46\columnwidth]{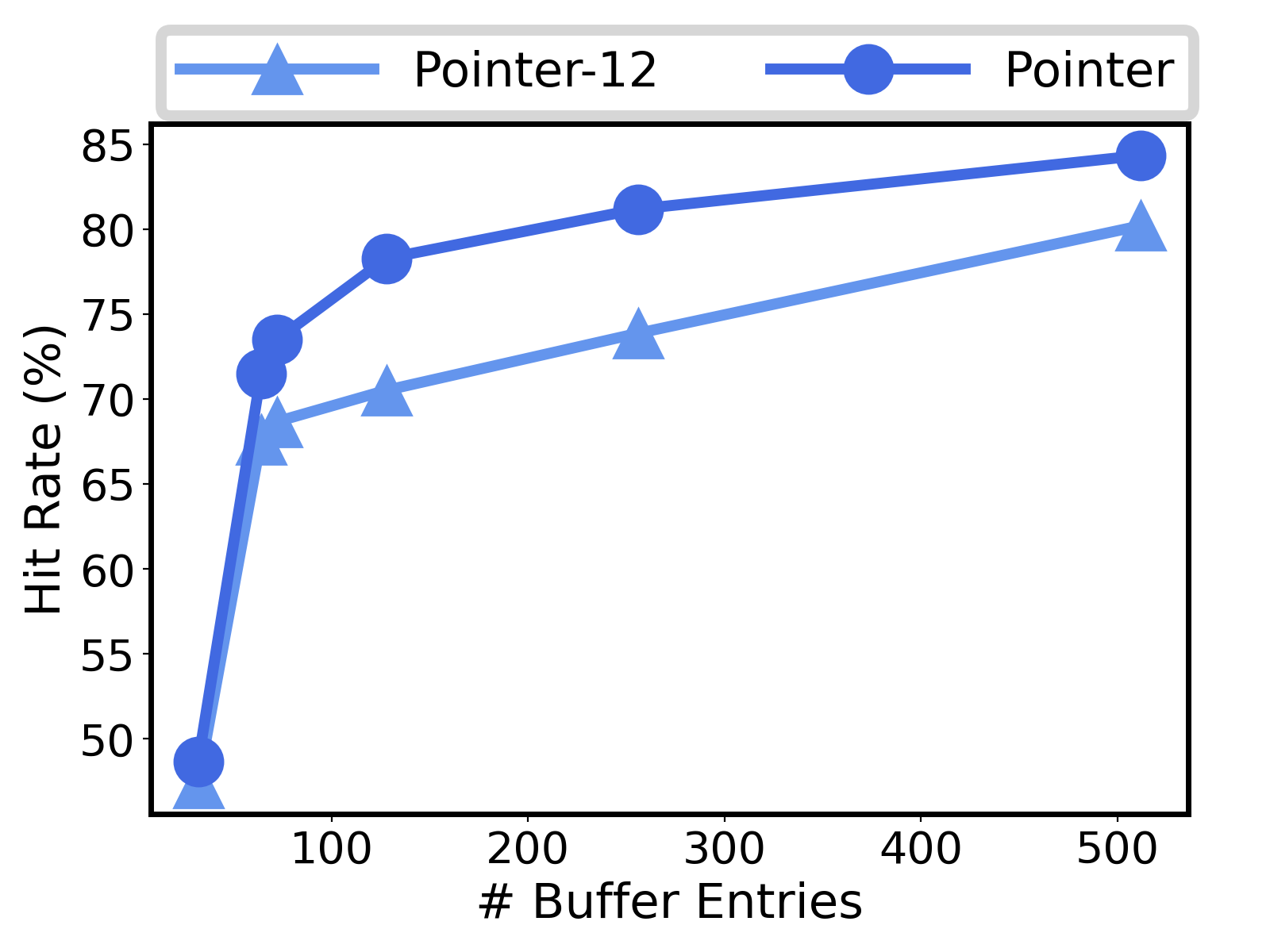}
    \label{cachehit1}
}
\subfigure[Hit rate in layer 2]{
    \centering
    \includegraphics[width=0.46\columnwidth]{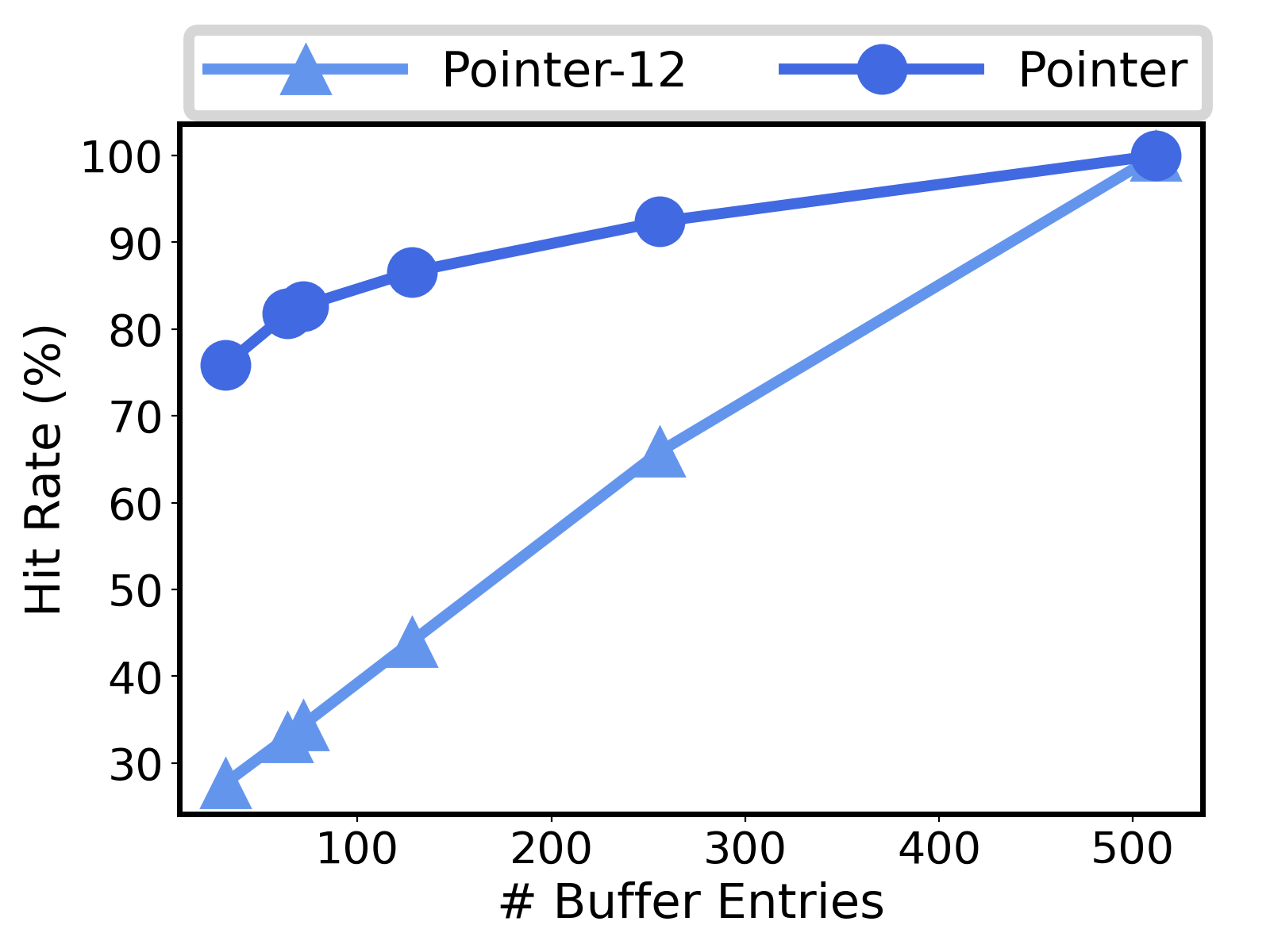}
    \label{cachehit2}
}
\vspace{-.13in}
\caption{On-chip buffer hit rate with different buffer sizes. No buffer for Pointer-1 so it is not shown.}
\label{cachehit}
\vspace{-.15in}
\end{figure}

The buffer size has a large impact on the performance. Fig.~\ref{perfsize} shows how the speedup values change with respect to the buffer size in Pointer-12 and Pointer. The buffer size has such performance impact since it directly affects the on-chip buffer hit rate. Here we further explore the correlation between on-chip buffer hit rate and buffer size in Fig.~\ref{cachehit1} and Fig.~\ref{cachehit2}.

For the first layer, as shown in Fig.~\ref{cachehit1}, because there are a larger number of input points to process in this first layer, the hit rate is relatively low when the buffer is small, lower than 50\% for both Pointer-12 and Pointer. With the increase of the buffer size, the hit rate also increases, but the hit rate of Pointer increases more significantly than Pointer-12 due to its better data locality.

For the second layer, as shown in Fig.~\ref{cachehit2}, the on-chip buffer hit rate of Pointer is always higher than Pointer-12 when the buffer size is smaller than 512. As the buffer size increases, the gap between Pointer and Pointer-12 becomes smaller. It is because the effect of the poor data locality is less obvious when given a larger buffer. When the buffer size reaches 512, the hit rate is 100\% because there are only 512 points in the input point cloud of layer 2. 

\section{Conclusion}

In this paper, we propose Pointer, a ReRAM-based point cloud accelerator with inter- and intra-layer optimizations. We first design a basic ReRAM-based accelerator to accelerate feature computation. Then we propose inter-layer coordination and intra-layer reordering to reduce DRAM access. Experiments show that Pointer outperforms MARS under different model sizes and buffer sizes.
Our proposed techniques may be transferred to other applications with irregular feature vector fetching such as graph neural network. 

\section*{Acknowledgement}
This work is partially supported by National Natural Science Foundation of China 92364102, and ACCESS – AI Chip Center for Emerging Smart Systems, sponsored by InnoHK funding, Hong Kong SAR. 

\bibliographystyle{ACM-Reference-Format}
\bibliography{refs}

\end{document}